%% file: ms.tex
\documentclass[usenatbib]{mn2e}

\usepackage{psfig,url,deluxetable}

\normalsize

\title[Search for fast radio bursts in three Parkes surveys]
{A search for highly dispersed fast radio bursts in three
Parkes multibeam surveys}
\author[Crawford et al.]
{F.~Crawford$^{1}$, 
A.~Rane$^{2}$, 
L.~Tran$^{1}$, 
K.~Rolph$^{1}$,
D.~R.~Lorimer$^{2,3}$ and 
J.~P.~Ridley$^{4}$\\
\\
$^1$Department of Physics and Astronomy, Franklin and
Marshall College, P.O. Box 3003, Lancaster, PA~17604, USA\\
$^2$Department of Physics, West Virginia University,
PO~Box~6315, Morgantown, WV~26506, USA\\
$^3$National Radio Astronomy Observatory, PO Box 2, Green
Bank, WV~24944, USA\\
$^4$Institute of Engineering, Murray State University, 
Murray, KY~42071, USA\\
}
\let\leqslant=\leq 
\date{Draft version created \today}
\begin{document}
\maketitle
\newcommand{\setthebls}{
}
\setthebls

\begin{abstract} 
We have searched three Parkes multibeam 1.4~GHz surveys for the
presence of fast radio bursts (FRBs) out to a dispersion measure (DM)
of 5000 pc cm$^{-3}$. These surveys originally targeted the Magellanic
Clouds (in two cases) and unidentified gamma-ray sources at
mid-Galactic latitudes (in the third case) for new radio pulsars. In
previous processing, none of these surveys were searched to such a
high DM limit. The surveys had a combined total of 719 hr of Parkes
multibeam on-sky time. One known FRB, 010724, was present in our data
and was detected in our analysis but no new FRBs were found. After
adding in the on-sky Parkes time from these three surveys to the
on-sky time (7512 hr) from the five Parkes surveys analysed by Rane et
al., all of which have now been searched to high DM limits, we improve
the constraint on the all-sky rate of FRBs above a fluence level of
3.8 Jy ms at 1.4 GHz to $R = 3.3^{+3.7}_{-2.2} \times 10^{3}$ events
per day per sky (at the 99\% confidence level).  Future Parkes surveys
that accumulate additional multibeam on-sky time (such as the ongoing
high-resolution Parkes survey of the LMC) can be combined with these
results to further constrain the all-sky FRB rate.
\end{abstract}


\section{INTRODUCTION}

In recent years a number of short-duration (millisecond) radio bursts
(``fast radio bursts'', or FRBs) have been detected by the Parkes,
Arecibo, and Green Bank radio telescopes in large-scale pulsar
surveys. These bursts have characteristics which indicate that they
are not of terrestrial origin and are likely of extragalactic
origin. The broadband dispersion characteristics observed for FRBs
very closely obey the cold plasma dispersion law in which the signal
delay is proportional to the inverse square of the observing frequency
(e.g., Lorimer \& Kramer 2005\nocite{lk05}). This is expected if the
signal originates from an astrophysical source (unlike, e.g., similar
signals detected in some surveys, such as Perytons, which have been
identified as terrestrial microwave interference; Petroff et
al. 2015b\nocite{pkb+15}). The FRBs detected to date also have
dispersion measures (DMs) significantly larger than what the Galactic
plasma content along the line of sight is likely to account for
\citep{cl02}. This fact, along with the recently proposed association
of FRB~150418 with an elliptical galaxy at redshift 0.5
\citep{kjb+16}, suggests that these bursts originate from very large
(cosmological) distances.  Note, however, that \citet{wb16} and
\citet{vrm+16} have called into question this association, and other
models have been proposed in which the high DM can be accounted for
more locally \citep[see, e.g.,][]{csp16}.

To date, only one of these bursts, FRB 121102 \citep{sch+14}, has been
observed to repeat \citep{ssh+16a,ssh+16b}, despite numerous efforts
and many hours spent trying to redetect FRBs in the same sky location
using the same observing system. With the possible exception of
FRB~150418, no FRBs have yet been localized to the point where
associations with known objects can be established. Thus, the physical
origin of FRBs remains uncertain, though the repeating nature of at
least a subset of FRBs indicates that some of them do not originate
from a cataclysmic event that destroys the source object. Models for
FRBs such as supergiant pulses emanating from magnetars in other
galaxies (e.g., Cordes \& Wasserman 2016\nocite{cw16}) are currently
favored.  For a recent overview and list of references to a variety of
proposed models for FRBs, see \citet{rlb+16} and \citet{kat16}.

The current tally of FRBs that have been detected and published is
presented in the Swinburne FRB Catalogue
\citep{pbj+16}.\footnote{http://www.astronomy.swin.edu.au/pulsar/frbcat/}
All but two of these FRBs were detected with the Parkes 64-m
telescope, and all but one have been detected at or near an observing
frequency of 1400 MHz. Efforts are now underway both to comb existing
pulsar survey data for FRBs that may have been missed in previous
analyses of the data and to detect FRBs as they occur using real-time
observing and detection systems. Examples of the latter include the
Parkes telescope at 1400 MHz \citep{pbb+15}, the ARTEMIS backend and
LOFAR array operating at a much lower frequency of 145 MHz
\citep{kca+15}, and the BURST project with the Molonglo Observatory
Synthesis Telescope which operates at an intermediate frequency of 843
MHz \citep{cfb+16}.

When the expected DM contribution from the Galaxy is removed using the
\citet{cl02} Galactic electron model, none of the FRBs detected to
date has an extragalactic DM contribution (DM excess) larger than
$\sim 1550$ pc cm$^{-3}$ \citep{cpk+16}.  \cite{zok+14} have shown
that there is a complicated non-linear relationship between the DM
contribution from the intergalactic medium (IGM) and
redshift. However, as seen in their Fig. 1, for small redshifts ($z
\la 3$), a linear approximation can be made in which $\sim 900$-1100
pc cm$^{-3}$ of DM is contributed per redshift unit. A DM excess of
$\sim 1550$ would then correspond to a redshift of $z \sim 1.5$
assuming that the IGM is the primary source of the dispersion.
However, significant local dispersion near the source or contributions
from the host galaxy could further boost the measured DM. Thus, if
FRBs beyond this redshift range are to be discovered, larger DMs must
be searched for burst signals.

We have searched three Parkes radio pulsar surveys to try and detect
putative bursts at very large DMs (up to a DM of 5000 pc
cm$^{-3}$). These three surveys were previously searched for FRBs in
the single-pulse search analysis done during the original data
processing, and in fact one of the surveys contains the first FRB
found, FRB~010724 \citep{lbm+07}. One of the other surveys has a known
Peryton present \citep{bbe+11}. However, none of the three surveys
have yet been searched out to high DMs ($> 1000$ pc cm$^{-3}$). All
three of the surveys targeted sky regions away from the Galactic plane
(all beams had Galactic latitudes $|b| > 5^{\circ}$; see Table
\ref{tbl-1}). This avoids foreground effects from the Galaxy that can
negatively affect the detectability of FRBs through increased pulse
scattering and sky temperature \citep{bb14,pvj+14}.

Below we describe each of the three surveys we searched and outline
our FRB search procedure. We then describe our results and the
subsequent constraints on the all-sky rate of FRBs.

\section{OVERVIEW OF THE SURVEYS}\label{sec:searches} 

The three surveys we have analysed were all conducted with the Parkes
radio telescope using the 13-beam multibeam receiver
\citep{swb+96}. Note that all but two of the FRBs detected and listed
to date in the Swinburne FRB Catalogue \citep{pbj+16} were detected
with this same observing system. Table \ref{tbl-1} describes the
observing parameters for each of these three surveys, which have a
cumulative multibeam on-sky time of 719 hr.

The first survey (``SMC'') was a deep search for pulsars in the
Magellanic Clouds. Both the Small (SMC) and Large (LMC)
Magellanic Clouds were searched with the same analog filterbank system
as was used in the highly successful Parkes Multibeam Pulsar Survey
\citep{mlc+01}. A total of 22 new pulsars were discovered in this
survey during several processing passes through the data
\citep{ckm+01,mfl+06,rcl+13}. The first FRB ever discovered, FRB
010724 \citep{lbm+07}, was also detected in this survey.  Prior to the
work described here, this survey had only been searched for periodic
signals and single pulses out to a maximum DM of 800 pc cm$^{-3}$. A
total of 488 hr of on-sky time was recorded in the survey.

The second survey (``EGU'') targeted 56 unidentified
mid-Galactic-latitude gamma-ray sources from the third EGRET catalog
\citep{hbb+99}.  The same observing system was used for this survey as
for the SMC survey described above, but with different integration and
sampling times (see Table \ref{tbl-1}). The results of the survey were
presented by \citet{crh+06}. The data were previously searched out to
DM = 1000 pc cm$^{-3}$, and six new pulsars were discovered. One of
these was PSR~J1614$-$2230, a binary system with a pulsar mass of
$1.97 \pm 0.04~M_{\odot}$ \citep{dpr+10}. A Peryton RFI burst signal
was also discovered in this survey \citep{bbe+11}. This survey
recorded a total of 135 hr of on-sky time.

The third survey (``PLMC'') is a new pulsar and transient survey of
the LMC which is sensitive to millisecond pulsars in the LMC
for the first time.  Like the two surveys above, it uses the Parkes
telescope and the multibeam receiver, but with the
Berkeley-Parkes-Swinburne Recorder (BPSR) digital backend
\citep{kjv+10}. This has a fast sampling capability and narrow
frequency channels (see Table \ref{tbl-1} for details), and 20\% of
the total survey data has been collected and processed so far
(corresponding to 96 hr of on-sky time).  The initial results from
this work were described by \citet{rcl+13}, where 3 new pulsars were
discovered. In this processing, the data were searched for pulsations
and single bursts out to DM = 500 pc cm$^{-3}$, but no new FRBs have
yet been detected in this survey.

\section{ANALYSIS}

In our re-analysis of these surveys, we searched the data for
impulsive signals at a much larger range of DMs than previously
searched.  We searched DMs ranging from 0 to 5000 pc cm$^{-3}$ with a
variable DM trial spacing that had a wider spacing at larger DMs. The
spacing was chosen so that the smearing introduced from a DM offset
would not significantly increase the DM smearing already present
within the finite frequency channels. Table \ref{tbl-1} lists the
number of DM trials used in each survey analysis.

Each dedispersed time series was searched for signals using a
widely-used single-pulse detection algorithm in the
SIGPROC\footnote{http://sigproc.sourceforge.net/} pulsar analysis
package.  This algorithm is described in detail by \citet{cm03} (see
also \citet{rlb+16} for a discussion) and uses a boxcar smoothing
technique to maintain sensitivity to pulses at a wide range of
time-scales. The boxcar filters were produced by averaging adjacent
time samples in 10 successive groups of two, yielding boxcar widths
ranging from 1 to 1024 time samples (see Table \ref{tbl-1} for the
sampling times used for the different surveys). The boxcar sample
aggregation was successively applied to each dedispersed time series,
with the highest resulting signal-to-noise ratio (S/N) signal from the
passes through the data being kept.  Only signals with a S/N $\ge 5$
were recorded. Note that this technique has been shown by \citet{kp15}
to reduce sensitivity to events which are offset from the boxcar
centers by as much as a factor of $\sqrt{2}$ (as compared to a
convolution of the time series with comparable boxcar filters). This
sensitivity reduction was taken into account in our estimate of the
all-sky FRB event rate.

A single-pulse diagnostic plot was produced for each beam (see Fig.
\ref{fig-1}). In this plot, a short, dispersed impulse would appear in
the DM vs. time plot as a signal at a non-zero DM that is localized in
both dimensions.  The size of the circles indicates the S/N.

No radio frequency interference (RFI) excision was performed prior to
the dedispersion and pulse search. However, RFI was cleaned in the
resulting single-pulse plots. If an RFI signal appeared at a DM of
zero (indicating a terrestrial signal), then samples at all DMs
corresponding to the time of that event were removed.  This technique
efficiently removed both non-dispersed broadband RFI and sporadic
narrowband RFI while maintaining the detectability of dispersed
broadband pulses (see Fig. \ref{fig-1}). After cleaning, both the
cleaned and uncleaned plots were checked by eye for indications of
dispersed pulse events.

\section{RESULTS AND DISCUSSION}

We discovered no new FRBs in the three surveys we searched out to a DM
of 5000 pc cm$^{-3}$. We did clearly detect both FRB~010724
\citep{lbm+07} in an SMC survey beam and the Peryton interference
signal present in several EGU survey beams \citep{bbe+11}. This signal
has been identified as a source of RFI \citep{pkb+15}, but since it
mimics some of the characteristics of FRBs it is a good test of our
single-pulse detection algorithm.

Both of these detections were made blindly (i.e., during the routine
analysis of the survey data). These detections are shown in Fig.
\ref{fig-1}, and these are the only FRB-type signals known to be
present in these surveys. Note that in addition to these two
single-burst events, a number of known pulsars were also detected
during the processing as single-pulse sources.

We used the null detection of any new FRBs in these three surveys we
analysed plus the results from five other Parkes multibeam 1.4 GHz
surveys that have been searched for FRBs (see Rane et
al. 2016\nocite{rlb+16}) to determine a new constraint on the all-sky
FRB rate.  These other large-scale Parkes surveys have all been
searched out to high DMs (at least a DM of 3000 pc cm$^{-3}$; see
Table 2 of Burke-Spolaor \& Bannister 2014\nocite{bb14}).

The five additional surveys we included in our FRB rate estimate are
listed below (see also Table 2 of \citet{rlb+16} which gives
additional details of each survey):

\begin{itemize} 

\item The high-time resolution universe south (HTRU-S) high-latitude
survey \citep{cpk+16}. A total of 2812 hr of on-sky time was recorded
in this survey and 9 new FRBs were discovered.

\item The HTRU-S intermediate-latitude survey \citep{pvj+14}. A total
of 1154 hr of on-sky time was recorded. No new FRBs were found.

\item The Swinburne Multibeam (SWMB) Survey \citep{bb14}. 925 hr total
was recorded and one new FRB was discovered.

\item The Parkes Multibeam Pulsar Survey (PMPS) \citep{mlc+01}. This
survey targeted low Galactic latitudes and had an on-sky integration
of 2115 hr. One new FRB was detected here.

\item The Parkes High-latitude (PH) Survey \citep{bjd+06}. 506 hr of
total on-sky time was recorded with no new FRBs found.

\end{itemize} 

We added the 7512 hr of time from the surveys above to the 719 hr from
our three surveys, and following the method of \citet{rlb+16}, we ran
a likelihood analysis to determine a statistically likely all-sky rate
of detectable FRBs. From this combined survey set, we find a rate of
$R = 3.3^{+3.7}_{-2.2} \times 10^{3}$ events per day per sky above a
fluence limit of 3.8 Jy ms at the 99\% confidence level. This is an
improvement over the \citet{rlb+16} limit of $R = 4.4^{+5.2}_{-3.1}
\times 10^{3}$ events per day above a 4 Jy ms fluence limit (99\%
confidence). Fig. \ref{fig-2} shows the likelihood function for both
the old and new all-sky rates.

Our derived FRB event rate above a uniform fluence threshold combines
the results of the rates determined individually from the 8 different
Parkes surveys, while also accounting for the different single-pulse
search processing methods and different telescope backends used in
these surveys (see Rane et al. 2016\nocite{rlb+16} for further
details). Other rate estimates that have been published from Parkes
observations have used only a single survey or subset of surveys
(e.g., the HTRU-S survey analysed by Champion et al.
2016\nocite{cpk+16} and Keane \& Petroff 2015\nocite{kp15}) which have
a smaller total on-sky time than the combined set of surveys that we
used. Given the large uncertainties in all of these rates (including ours),
they are all compatible with each other. However, our rate is an
improvement on the recent \citet{rlb+16} rate estimate since we have
included the on-sky time from 3 more surveys to their analysis and
detected no additional FRBs.  We note that the PMPS was conducted at
low Galactic latitudes ($|b| < 5^{\circ}$), and Galactic-plane effects
may significantly influence detectability of any FRBs present and
hence can affect underlying FRB rate estimates.

The inclusion of this information in the future analysis of other
Parkes multibeam surveys (such as the complete PLMC survey, of which
only 20\% has been observed and processed; Ridley et
al. 2013\nocite{rcl+13}) will help further constrain the all-sky FRB
rate.

\section{CONCLUSIONS}

We have analysed three Parkes multibeam surveys for FRBs at a range of
DMs extending out to 5000 pc cm$^{-3}$, a much higher DM limit than
what was previously searched in these surveys. We detected one known
FRB and one known Peryton interference signal that were present in
these surveys, but found no new FRBs. We used the 719 hr of multibeam
on-sky time from our three surveys and the 7512 hr of on-sky time from
five other large-scale Parkes multibeam surveys searched out to high
DMs (at least 3000 pc cm$^{-3}$) to improve the constraint on the FRB
all-sky rate.  We determine a rate of $R = 3.3^{+3.7}_{-2.2} \times
10^{3}$ events per day per sky above a fluence limit of 3.8 Jy ms at
the 99\% confidence level. Results from future Parkes surveys will be
able to be combined with these results to further constrain the
underlying FRB rate.

\section*{ACKNOWLEDGMENTS}

The Parkes radio telescope is part of the Australia Telescope which is
funded by the Commonwealth of Australia for operation as a National
Facility managed by CSIRO. Work at Franklin \& Marshall College was
partially supported by the Hackman scholarship fund. FC thanks the
McGill Space Institute for hospitality during the completion of this
manuscript.

\begin{figure*}
\centerline{\psfig{figure=lorimer.ps,width=5.0in,angle=270}}
\centerline{\psfig{figure=lorimer_cleaned.ps,width=5.0in,angle=270}}
\centerline{\psfig{figure=peryton.ps,width=5.0in,angle=270}}
\centerline{\psfig{figure=peryton_cleaned.ps,width=5.0in,angle=270}}
\caption{Single-pulse detections of two known burst signals present in
the surveys we analysed: FRB~010724 \citep{lbm+07} from the SMC survey
(shown in the top two panels), and a Peryton RFI signal \citep{bbe+11}
from the EGU survey (shown in the bottom two panels). Each pair of
panels shows DM vs. time prior to RFI cleaning (top panel) and after
cleaning (bottom panel) for that particular source. The symbol size
indicates signal strength.  The cleaning process removes undispersed
broadband terrestrial RFI (clustered at DM = 0 throughout the
integration) and narrowband RFI (occasional thin vertical signals)
while preserving broadband dispersed signals.  Both signals are
clearly detected in the diagnostic plots.\label{fig-1}}
\end{figure*}

\begin{figure*}
\centerline{\psfig{figure=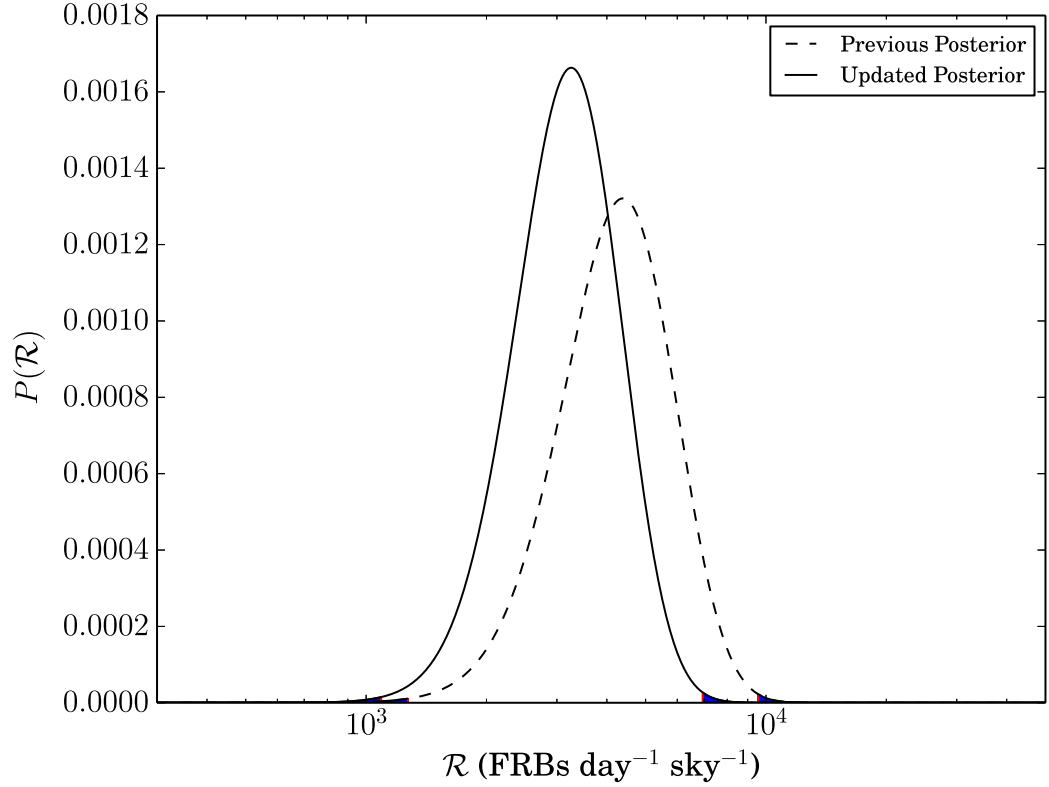,width=6.0in,angle=0}}
\caption{The posterior probability density function of the event rate
of Parkes-detectable FRBs, determined from the 5 Parkes surveys
(totaling 7512 hr) analysed by \citet{rlb+16} (dashed curve) and from
the addition of the 3 Parkes surveys (totaling 719 hr) described in
this paper (solid curve). See also Table 2 of \citet{rlb+16}.  All
surveys were searched to high DMs (at least 3000 pc cm$^{-3}$). The
rate analysis considered the different single-pulse search processing
methods and observing backends used in the different surveys.  The
resulting new all-sky FRB rate is $R = 3.3^{+3.7}_{-2.2} \times
10^{3}$ events per day per sky above a fluence limit of 3.8 Jy ms at
the 99\% confidence level.\label{fig-2}}
\end{figure*}

\onecolumn

\input{tab1}  

\end{document}

%% file: tab1.tex
\begin{deluxetable}{lccc}
\tabletypesize{\footnotesize}
\tablecaption{Summary of Three Parkes Surveys Searched\label{tbl-1}}
\tablewidth{0pt}
\tablehead{
\colhead{Survey} &
\colhead{SMC} &
\colhead{EGU} &
\colhead{PLMC} 
}
\startdata 
Galactic Latitude Range                                   & $|b| \sim 45^{\circ}$ (SMC); & $5^{\circ} < |b| < 73^{\circ}$     & $|b| \sim 53^{\circ}$ (LMC) \\ 
                                                          & $|b| \sim 53^{\circ}$ (LMC)  &                                    &                             \\
Total Number of Survey Beams                              & 2717\tablenotemark{a}        & 3016                               & 520\tablenotemark{b}        \\
Integration Time Per Pointing (s)                         & 8400                         & 2100                               & 8600                        \\
Total On-sky Survey Time (hr)                             & 488                          & 135                                & 96\tablenotemark{b}         \\
Sampling Time (ms)                                        & 1.000                        & 0.125                              & 0.512\tablenotemark{c}      \\
Number of Frequency Channels                              & 96                           & 96                                 & 870\tablenotemark{d}        \\
Observing Bandwidth (MHz)                                 & 288                          & 288                                & 340\tablenotemark{d}        \\
Center Observing Frequency (MHz)                          & 1374                         & 1374                               & 1352\tablenotemark{d}       \\
Max. Galactic DM Contribution (pc cm$^{-3}$)\tablenotemark{e}  & $\sim 50$               & $\sim 500$\tablenotemark{f}        & $\sim 50$                   \\
Original Max. DM Searched (pc cm$^{-3}$)\tablenotemark{g} & 800                          & 1000                               & 500                         \\
New Max. DM Searched (pc cm$^{-3}$)                       & 5000                         & 5000                               & 5000                        \\
Number of Trial DMs in New Search                         & 256                          & 371                                & 1431                        \\
Known Burst Signals Detected                              & 1 FRB\tablenotemark{h}       & 1 Peryton\tablenotemark{i}         & $-$                         \\
Survey References                                         & \citet{mfl+06};              & \citet{crh+06}                     & \citet{rcl+13}              \\
                                                          & \citet{rcl+13}               &                                    &                            
\enddata

\tablecomments{All three surveys used the 13-beam multibeam receiver
\citep{swb+96} on the Parkes 64-m telescope, and all surveys were
conducted at 1.4 GHz. All three surveys therefore had the same beam
size and instantaneous sensitivity as other large-scale Parkes surveys
recently searched for FRBs.}

\tablenotetext{a}{Our analysis used 2756 beams (2717 original survey
beams plus 39 unique extra beams that were not used in the
\citet{mfl+06} survey grid).}

\tablenotetext{b}{This corresponds to the first 20\% of the total
survey coverage, which is the fraction of the survey that has been
observed (and processed) to date.}

\tablenotetext{c}{For the analysis here, the raw time samples were
aggregated into groups of 8 to create an effective sampling time of
0.512 ms from the native 0.064 ms sampling at the telescope recorder.}

\tablenotetext{d}{The BPSR data recorder used at the Parkes telescope
has 400 MHz of bandwidth split into 1024 channels with a 1382 MHz
center observing frequency \citep{kjv+10}. However, the receiver is
not sensitive to the top 60 MHz of the band, which is blanked during
the data analysis. The table therefore shows the effective values with
this 60 MHz band removed.}

\tablenotetext{e}{Maximum Galactic DM contribution estimated from the
NE2001 Galactic electron model \citep{cl02} for all survey lines of
sight.}

\tablenotetext{f}{The expected maximum Galactic DM along the line of
sight is $\la 100$ pc cm$^{-3}$ for more than half of the target
sources in this survey, and no lines of sight have an expected maximum
Galactic DM greater than 500 pc cm$^{-3}$.}

\tablenotetext{g}{Maximum DM searched for pulsars and impulsive
signals in the original survey analysis.} 

\tablenotetext{h}{FRB 010724 was discovered by \citet{lbm+07} in this
survey with DM = 375 pc cm$^{-3}$ (see Fig.~\ref{fig-1}). This signal
was detected in our analysis.}

\tablenotetext{i}{One Peryton was discovered by \citet{bbe+11} which
had a fitted DM $\sim 375$ pc cm$^{-3}$. This signal was detected in
our analysis (see Fig.~\ref{fig-1}), but it has been
determined to be terrestrial in origin \citep{pkb+15}.}

\end{deluxetable}

%% file: ms.bbl
\begin{thebibliography}
\bibliography{}

\bibitem[Burgay et al.(2006)]{bjd+06} Burgay, M., Joshi,
B.~C., D'Amico, N., et al.\ 2006, MNRAS, 368, 283

\bibitem[Burgay et al.(2013)]{bkl+13} Burgay, M., Keith, 
M.~J., Lorimer, D.~R., et al.\ 2013, MNRAS, 429, 579 

\bibitem[Burke-Spolaor et al.(2011)]{bbe+11} Burke-Spolaor, S.,
Bailes, M., Ekers, R., et al.\ 2011, ApJ, 727, 18

\bibitem[Burke-Spolaor \& Bannister(2014)]{bb14} Burke-Spolaor, S., \&
Bannister, K.~W.\ 2014, ApJ, 792, 19

\bibitem[Caleb et al.(2016)]{cfb+16} Caleb, M., Flynn, C., 
Bailes, M., et al.\ 2016, MNRAS, 458, 718 

\bibitem[Champion et al.(2016)]{cpk+16} Champion, D.~J., Petroff, E.,
Kramer, M., et al.\ 2016, MNRAS, 460, L30 

\bibitem[Connor et al.(2016)]{csp16} 
Connor, L., Sievers, J., \& Pen, U.-L.\ 2016, MNRAS, 458, L19 

\bibitem[Cordes \& Lazio(2002)]{cl02} Cordes, J. M. \& Lazio,
T. J. W. 2002, astro-ph/0207156

\bibitem[Cordes \& McLaughlin(2003)]{cm03} Cordes, J.~M., \&
McLaughlin, M.~A.\ 2003, ApJ, 596, 1142

\bibitem[Cordes \& Wasserman(2016)]{cw16} Cordes, J.~M., \& Wasserman,
I.\ 2016, MNRAS, 457, 232

\bibitem[Crawford et al.(2001)]{ckm+01} Crawford, F., Kaspi, V.~M.,
Manchester, R.~N., et al.\ 2001, ApJ, 553, 367

\bibitem[Crawford et al.(2006)]{crh+06} Crawford, F., Roberts,
M.~S.~E., Hessels, J.~W.~T., et al.\ 2006, ApJ, 652, 1499

\bibitem[Demorest et al.(2010)]{dpr+10} Demorest, P.~B., Pennucci, T.,
Ransom, S.~M., et al..\ 2010, Nature, 467, 1081

\bibitem[Hartman et al.(1999)]{hbb+99} Hartman, R.~C.,
Bertsch, D.~L., Bloom, S.~D., et al.\ 1999, ApJS, 123, 79

\bibitem[Karastergiou et al.(2015)]{kca+15} Karastergiou, A.,
Chennamangalam, J., Armour, W., et al.\ 2015, MNRAS, 452, 1254

\bibitem[Katz(2016)]{kat16} Katz, J.~I.\ 2016, Modern Physics Letters A, 31, 
1630013  

\bibitem[Keane et al.(2016)]{kjb+16} Keane, E.~F., Johnston, S.,
Bhandari, S., et al.\ 2016, Nature, 530, 453

\bibitem[Keane \& Petroff(2015)]{kp15} Keane, E.~F., \& Petroff, 
E.\ 2015, MNRAS, 447, 2852 

\bibitem[Keane et al.(2012)]{ksk+12} Keane, E.~F., Stappers,
B.~W., Kramer, M., \& Lyne, A.~G.\ 2012, MNRAS, 425, L71

\bibitem[Keith et al.(2010)]{kjv+10} Keith, M.~J., Jameson,
A., van Straten, W., et al.\ 2010, MNRAS, 409, 619

\bibitem[Lorimer \& Kramer(2005)]{lk05} Lorimer, D. R., \& Kramer,
M. 2005, Handbook of Pulsar Astronomy (Cambridge: Cambridge University
Press)

\bibitem[Lorimer et al.(2007)]{lbm+07} Lorimer, D.~R., Bailes, M.,
McLaughlin, M.~A., et al.\ 2007, Science, 318, 777

\bibitem[Manchester et al.(2001)]{mlc+01} Manchester, R.~N.,
Lyne, A.~G., Camilo, F., et al.\ 2001, MNRAS, 328, 17

\bibitem[Manchester et al.(2006)]{mfl+06} Manchester, R.~N., Fan, G.,
Lyne, A.~G., et al..\ 2006, ApJ, 649, 235

\bibitem[Petroff et al.(2014)]{pvj+14} Petroff, E., van Straten, W.,
Johnston, S., et al.\ 2014, ApJ, 789, L26

\bibitem[Petroff et al.(2015a)]{pbb+15} Petroff, E., Bailes, M., Barr,
E.~D., et al.\ 2015a, MNRAS, 447, 246

\bibitem[Petroff et al.(2015b)]{pkb+15} Petroff, E., Keane, E.~F.,
Barr, E.~D., et al.\ 2015b, MNRAS, 451, 3933

\bibitem[Petroff et al.(2016)]{pbj+16} Petroff, E., Barr, E.~D.,
Jameson, A., et al.\ 2016, Publications of the
Astronomical Society of Australia, submitted, arXiv:1601.03547

\bibitem[Rane et al.(2016)]{rlb+16} Rane, A., Lorimer, D.~R., Bates,
S.~D., et al.\ 2016, MNRAS, 455, 2207

\bibitem[Ridley et al.(2013)]{rcl+13} Ridley, J.~P., Crawford,
F., Lorimer, D.~R., et al.\ 2013, MNRAS, 433, 138

\bibitem[Scholz et al.(2016)]{ssh+16b} Scholz, P., Spitler, L.~G.,
Hessels, J.~W.~T., et al.\ 2016, ApJ, submitted, arXiv:1603.08880

\bibitem[Spitler et al.(2014)]{sch+14} Spitler, L.~G., Cordes, J.~M.,
Hessels, J.~W.~T., et al.\ 2014, ApJ, 790, 101

\bibitem[Spitler et al.(2016)]{ssh+16a} Spitler, L.~G., Scholz, P.,
Hessels, J.~W.~T., et al.\ 2016, Nature, 531, 202 

\bibitem[Staveley-Smith et al.(1996)]{swb+96} Staveley-Smith, L.,
Wilson, W.~E., Bird, T.~S., et al.\ 1996, Publications of the
Astronomical Society of Australia, 13, 243

\bibitem[Vedantham et al.(2016)]{vrm+16} Vedantham, H.~K., Ravi, V., Mooley, K., et al.\ 2016, ApJ, submitted, arXiv:1603.04421

\bibitem[Williams \& Berger(2016)]{wb16} Williams, P.~K.~G., \&
Berger, E.\ 2016, arXiv:1602.08434

\bibitem[Zheng et al.(2014)]{zok+14} Zheng, Z., Ofek, E.~O., Kulkarni,
S.~R., et al.\ 2014, ApJ, 797, 71


\end{thebibliography}
